\documentclass[sigconf]{acmart}
\usepackage{algorithm}
\usepackage{algpseudocode}
\usepackage{amsmath}
\usepackage{booktabs}
\usepackage[table]{xcolor}
\usepackage{makecell}

\usepackage{amsmath,amsfonts}
\usepackage{graphicx, subfigure, verbatim, url,pifont}
\usepackage{stmaryrd}
 \usepackage{multirow}
 \usepackage{booktabs}
\usepackage{makecell}

\AtBeginDocument{%
  }

\setcopyright{acmlicensed}
\copyrightyear{2026}
\acmYear{2026}
\acmDOI{XXXXXXX.XXXXXXX}
\acmConference[Conference acronym 'XX]{Make sure to enter the correct
  conference title from your rights confirmation email}{June 03--05,
  2026}{Woodstock, NY}
\acmISBN{978-1-4503-XXXX-X/2026/06}

\begin{document}

\title{MSAO: Adaptive Modality Sparsity-Aware Offloading with Edge-Cloud Collaboration for Efficient Multimodal LLM Inference}


\author{Zheming Yang\textsuperscript{\rm 1},
    Qi Guo\textsuperscript{\rm 1,2},
    Jun Wan\textsuperscript{\rm 3},
    Jiarui Ruan\textsuperscript{\rm 4},
    Yunqing Hu\textsuperscript{\rm 1,2},
    Chang Zhao\textsuperscript{\rm 1,2}, and
    Xiangyang Li\textsuperscript{\rm 1}\\}
\affiliation{%
  \textsuperscript{\rm 1}\institution{Institute of Computing Technology, Chinese Academy of Sciences}
\textsuperscript{\rm 2}\institution{University of Chinese Academy of Sciences}
\textsuperscript{\rm 3}\institution{University of Science and Technology of China}
 \textsuperscript{\rm 4}\institution{University of Illinois at Urbana-Champaign}
 \textcolor{white}{ \country{China}}
  }

\renewcommand{\shortauthors}{Trovato et al.}

\begin{abstract}
Multimodal large language models (MLLMs) enable powerful cross-modal reasoning capabilities but impose substantial computational and latency burdens, posing critical challenges for deployment on resource-constrained edge devices. In this paper, we propose MSAO, an adaptive modality sparsity-aware offloading framework with edge-cloud collaboration for efficient MLLM Inference. First, a lightweight heterogeneous modality-aware via fine-grained sparsity module performs spatial-temporal-modal joint analysis to compute the Modality Activation Sparsity (MAS) metric, which quantifies the necessity of each modality with minimal computational overhead. Second, an adaptive speculative edge-cloud collaborative offloading mechanism dynamically schedules workloads between edge and cloud based on the derived MAS scores and real-time system states, leveraging confidence-guided speculative execution to hide communication latency. Extensive experiments on VQAv2 and MMBench benchmarks demonstrate that MSAO achieves a 30\% reduction in end-to-end latency and 30\%--65\% decrease in resource overhead, while delivering a throughput improvement of 1.5$\times$ to 2.3$\times$ compared to traditional approaches, all without compromising competitive accuracy.


\end{abstract}




\maketitle

\section{Introduction}

\begin{figure}[t]
  \centering
   \includegraphics[width=1\columnwidth]{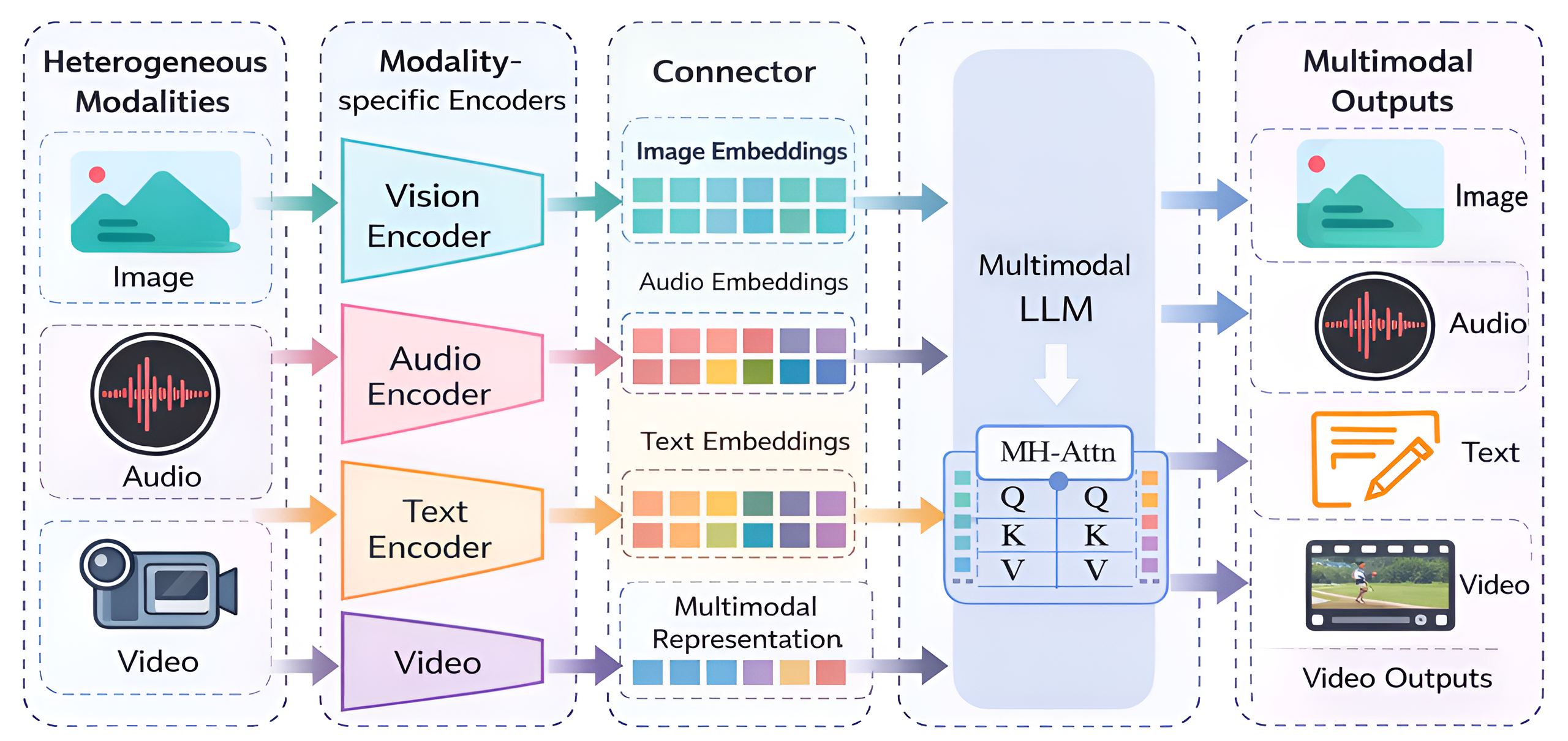}
  \caption{The overview of MLLM inference. Heterogeneous inputs (image, video, audio, text) are encoded separately, aligned into a unified token space, and processed by the LLM backbone to generate responses. }
  \label{figure1}
\end{figure}

Multimodal large language models (MLLMs) have recently demonstrated remarkable capabilities in integrating visual, auditory, and textual information, enabling unified reasoning across heterogeneous data sources \cite{wu2023multimodal111}, as illustrated in Figure~\ref{figure1}. These models are increasingly deployed in latency-sensitive applications ranging from intelligent assistants and autonomous systems to real-time perception pipelines \cite{koh2023generating222, 393939}. However, their impressive performance comes at substantial computational and memory costs \cite{lin2025boosting333}, as the joint processing of multiple modalities, each with distinct spatial, temporal, and semantic characteristics, exponentially increases the inference burden.  Cloud-based solutions suffer from excessive transmission latency and bandwidth contention \cite{zhang2025gsmm444}. Conversely, edge-only approaches offer lower response time and better data locality but are fundamentally constrained by limited computing capacity \cite{yao2025efficient555}, leading to degraded performance on complex multimodal tasks that require deep cross-modal reasoning \cite{zheng2025review666}.

To overcome these limitations, recent research has explored edge–cloud collaborative inference \cite{yang2025ec2moe777,wang2024cloud888}, where lightweight preprocessing is performed on edge devices while computation-intensive reasoning is offloaded to the cloud. Such hybrid solutions balance efficiency and accuracy by distributing workloads across heterogeneous resources \cite{zhang2024mm515151}. However, existing collaborative frameworks typically adopt uniform offloading policies that treat all modalities equally, overlooking the intrinsic heterogeneity across different input types \cite{yang2024perllm101010}. In practice, visual inputs (images and videos) often contain substantial spatial and temporal redundancy, such as irrelevant backgrounds or static scenes, that can be aggressively compressed without compromising task performance\cite{jin2025efficient454545, DBLP505050}. In contrast, textual and audio inputs may carry semantically dense information that requires high-fidelity processing \cite{miao2025towards444444}. Moreover, different modalities exhibit varying sensitivity to inference latency and bandwidth constraints. Ignoring these differences leads to suboptimal workload allocation, inefficient resource utilization, and inconsistent user experience \cite{yi2025enhancing121212}.  First, uniform offloading policies may waste bandwidth by transmitting redundant visual information that could have been processed or filtered on the edge \cite{hsia2024mad414141,jin2024adaptive424242}. Second, they may unnecessarily burden the edge device with computationally intensive modalities that would be better handled by the cloud, causing increased latency \cite{chen2024data434343}.

In this paper, we present MSAO, an adaptive modality sparsity-aware offloading framework with edge-cloud collaboration for efficient MLLM Inference. It introduces fine-grained modality sparsity analysis into the offloading decision process, enabling adaptive workload scheduling that balances inference quality, latency, and resource consumption. The main contributions are as follows:

\begin{itemize}
\item We develop a lightweight heterogeneous modality-aware via fine-grained sparsity module that performs spatial-temporal-modal joint analysis to compute the Modality Activation Sparsity (MAS) metric. This module quantifies the necessity of each modality with minimal computational overhead, enabling efficient identification of redundant information across image, video, and text inputs on resource-constrained edge devices.

\item We propose an adaptive speculative edge-cloud collaborative offloading mechanism that dynamically schedules workloads between edge and cloud based on the derived MAS scores and real-time system states. By leveraging confidence-guided speculative execution, this mechanism effectively hides communication latency behind computation, achieving near-optimal overlap between edge draft generation and cloud verification.

\item The experimental results demonstrate that MSAO achieves a 30\% reduction in end-to-end latency and a 30\%--65\% decrease in resource overhead, while delivering a throughput improvement of 1.5$\times$ to 2.3$\times$ compared to traditional approaches, all without compromising competitive accuracy.
\end{itemize}


    
    

\section{Related Work}

Some research has focused on optimizing MLLM inference in purely cloud-based or purely edge-based environments. Cloud-centric efforts include adaptive scheduling for edge computing \cite{10.1145/3774908}, compositional reasoning with structured task relations \cite{11207619}, and recursive offloading in multi-tier networks \cite{11293838}. Edge-only approaches explore the feasibility of deploying LLMs directly on edge devices \cite{zhang2024beyond}, and heterogeneity-aware distributed inference \cite{hu2025adaptlink}. To overcome the computational constraints of edge devices, several works have proposed speculative decoding and sparsity-aware acceleration techniques. SLED \cite{li2025sled} presents a speculative LLM decoding framework for efficient edge serving, enabling faster token generation through draft-then-verify mechanisms. SparCIM \cite{xu2025sparcim} leverages contextual and unstructured bit sparsity to accelerate LLM inference on heterogeneous compute-in-memory accelerators. ExeGPT \cite{oh2024exegpt} introduces constraint-aware resource scheduling for LLM inference, optimizing resource allocation under latency and power constraints. Recursive offloading for LLM serving in multi-tier networks \cite{11293838} proposes hierarchical offloading strategies that recursively partition inference tasks across network tiers. SpaceServe \cite{zhang2025spaceserve} explores spatial multiplexing of complementary encoders and decoders for multimodal LLMs, improving throughput by parallelizing different components of the model. While these approaches have made significant strides, cloud-only solutions suffer from high communication latency, while edge-only methods are constrained by limited model capacity.

Recent advances in edge-cloud collaborative inference have sought to balance the computational demands of MLLMs with the resource constraints of edge devices. Various approaches have been proposed, including semantically enhanced offloading for object detection \cite{Adaptive_Guidance2025}, scene-aware collaborative frameworks for industrial inspection \cite{SAEC}, and adaptive collaboration for visual detection \cite{AIVD}. Beyond industrial applications, researchers have explored cloud-edge architectures for driver assistance systems \cite{47}, collaborative inference for enhanced reasoning \cite{11086409}, task-oriented feature compression \cite{11222951}, decoupled video inference offloading \cite{yuan2026video}, cloud-edge adaptation for mobile networking \cite{50}, historical interaction reuse for on-device inference \cite{ding2024enhancing}, and joint offloading with model caching \cite{11154975}. Despite these advances, existing frameworks predominantly adopt uniform offloading policies that treat all modalities equally, overlooking the intrinsic heterogeneity across different input types. Visual inputs often contain spatial and temporal redundancy that can be aggressively compressed, while textual and audio inputs require high-fidelity processing.

\section{Preliminaries}
\label{sec:format}

\textbf{Multimodal LLM.} MLLMs extend text-only LLMs by integrating heterogeneous modalities such as images, audio, and video into a unified reasoning framework \cite{yin2024survey202020}. Formally, given an input set of modalities ${x_t, x_v, x_a}$ corresponding to text, vision, and audio, the model extracts modality-specific features via dedicated encoders $f_m(\cdot)$:
\begin{equation}
h_m = f_m(x_m), \quad m \in {t, v, a}.
\end{equation}
These features are then projected into a shared latent space through a learnable projection function $g(\cdot)$:
\begin{equation}
z_m = g(h_m), \quad m \in {t, v, a}.
\end{equation}
The unified embeddings are subsequently fed into a transformer-based backbone for cross-modal reasoning \cite{ge2024worldgpt464646, hu2024bliva474747}. Notably, different modalities contribute vastly different numbers of tokens; an image may yield hundreds of visual tokens while a text query contributes only a few, leading to significant heterogeneity in computational cost \cite{dong2025coef484848,wang2024multimodal494949}. This modality-aware heterogeneity motivates adaptive offloading strategies that exploit such variations to improve inference efficiency.

\textbf{Edge-Cloud Architecture.} Edge-cloud architectures provide a hierarchical computing paradigm that distributes inference workloads across resource-constrained edge devices and powerful cloud servers, balancing latency, energy efficiency, and computational capacity \cite{yang2021intelligent212121,yang2023javp363636}. Typically, the edge handles lightweight preprocessing and small-scale execution, while the cloud undertakes large-model inference and complex reasoning \cite{tong2016hierarchical131313,373737}. Based on factors such as modality composition, network bandwidth, and edge resource availability, the system dynamically decides whether to execute locally, offload fully to the cloud, or adopt a hybrid approach \cite{383838,404040}. This collaborative design mitigates the limitations of standalone edge inference while reducing the dependency on purely cloud-centric solutions, making it especially suitable for real-time multimodal applications under dynamic conditions.




\section{The Proposed MSAO Framework}
\label{sec:method}

\begin{figure}[t]
  \centering
 \includegraphics[width=1\columnwidth]{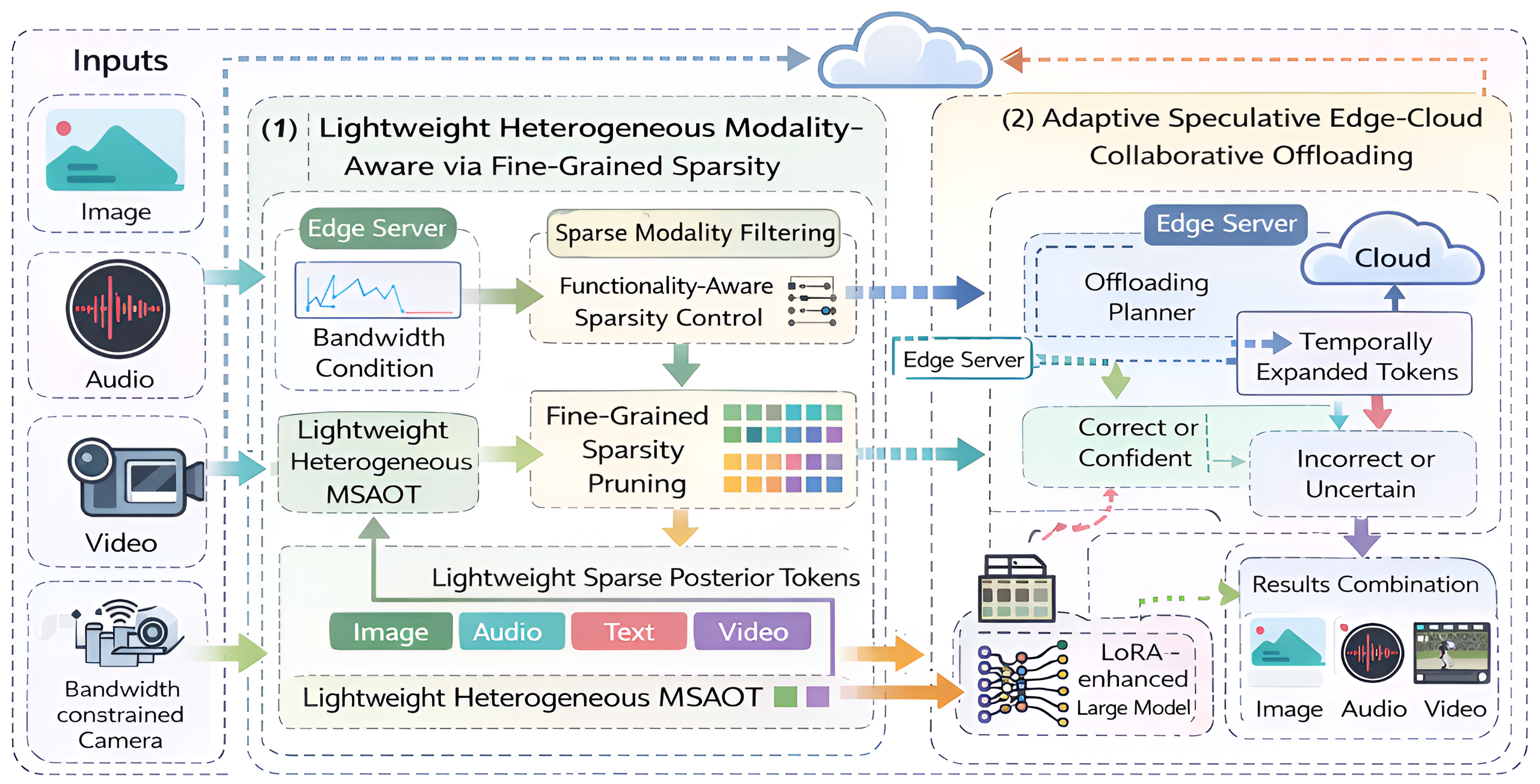}
  \caption{The overview of the proposed MSAO framework.}
  \label{figure2}
  \vspace{-2mm}
\end{figure}

We propose MSAO, an adaptive modality sparsity-aware offloading framework designed to enhance the MLLM inference efficiency by edge–cloud collaboration, as illustrated in Figure~\ref{figure2}. It primarily consists of two parts: (1) \textit{Lightweight Heterogeneous Modality-Aware via Fine-Grained Sparsity}, and (2) \textit{Adaptive Speculative Edge-Cloud Collaborative Offloading}. The core idea of MSAO is to leverage intrinsic modality sparsity as a guiding signal for adaptive offloading decisions. Unlike traditional approaches that treat all modalities uniformly or rely on heuristic-based compression, MSAO first identifies which parts of the multimodal input are critical for task completion and which are redundant, using a lightweight probing network with minimal computational overhead. Based on this fine-grained sparsity analysis, the framework then employs a confidence-guided speculative execution strategy that adaptively determines whether to execute locally on the edge or offload to the cloud and when to invoke speculative decoding to hide communication latency. By jointly optimizing modality selection, workload partitioning, and execution timing, MSAO achieves a principled tradeoff between inference accuracy, end-to-end latency, and resource consumption.

\subsection{Lightweight Heterogeneous Modality-Aware via Fine-Grained Sparsity}

To enable efficient edge-cloud collaborative inference for MLLMs, it is essential to first identify the computational redundancy inherent in the input data. Instead of directly processing full multimodal inputs with the heavy MLLM backbone, we propose a lightweight prefix probing network that performs \textit{fine-grained sparsity modeling} across spatial, temporal, and modal dimensions. This module quantifies the necessity of each modality for the given inference task, laying the foundation for subsequent adaptive offloading decisions.

\subsubsection{Spatial Sparsity}

For visual modalities (images and video frames), a significant portion of the input often consists of irrelevant backgrounds that contribute little to the final response. Leveraging the observation that early layers in vision encoders capture spatial structures with minimal computational overhead, we design a lightweight spatial probing head attached to the intermediate feature maps of the vision encoder.

Let $\mathbf{F}^{(l)} \in \mathbb{R}^{H \times W \times C}$ denote the feature map at layer $l$ of the vision encoder. We apply a lightweight convolutional predictor to generate a spatial importance map $\mathbf{M}_{\text{spatial}} \in \mathbb{R}^{H \times W}$:

\begin{equation}
\mathbf{M}_{\text{spatial}} = \sigma\left( \text{Conv}_{1\times1}\left( \text{AvgPool}(\mathbf{F}^{(l)}) \right) \right),
\end{equation}
where $\sigma(\cdot)$ denotes the sigmoid activation function. Each entry in $\mathbf{M}_{\text{spatial}}$ represents the importance of the corresponding image patch. We then define the \textit{spatial sparsity ratio} $\rho_{\text{spatial}}$ as the proportion of patches whose importance falls below a threshold $\tau_s$:

\begin{equation}
\rho_{\text{spatial}} = \frac{|\{ (i,j) \mid \mathbf{M}_{\text{spatial}}^{(i,j)} < \tau_s \}|}{H \times W}.
\end{equation}

Patches with low importance are identified as non-critical backgrounds and can be aggressively pruned or coarsely compressed in downstream processing.

\subsubsection{Temporal Sparsity}

For video inputs, consecutive frames often exhibit substantial temporal redundancy, especially in static scenes. To capture this, we introduce a temporal probing mechanism that computes the similarity between feature representations of adjacent frames with negligible overhead. Given the feature representations $\mathbf{f}_t$ and $\mathbf{f}_{t-1}$ of frames $t$ and $t-1$ extracted from the early layers of the vision encoder, we compute their hash-based similarity using the following locality-sensitive hashing (LSH) approximation:

\begin{equation}
\text{sim}_t = \frac{1}{K} \sum_{k=1}^{K} \mathbb{1}\left[ h_k(\mathbf{f}_t) = h_k(\mathbf{f}_{t-1}) \right],
\end{equation}
where $h_k(\cdot)$ denotes the $k$-th hash function, and $K$ is the total number of hash functions. The temporal redundancy score $\gamma_t$ for frame $t$ is defined as $\gamma_t = 1 - \text{sim}_t$. A low $\gamma_t$ indicates high redundancy with the previous frame, suggesting that frame $t$ can be safely subsampled or represented using only the residual difference.

\subsubsection{Modal Sparsity}

Different modalities contribute unequally to the final reasoning result depending on the input prompt. For instance, a text-based question about the color of an object may heavily rely on visual information, while a question about audio content may render visual input largely irrelevant. To assess modal importance, we design a lightweight cross-modal probing network that estimates the \textit{information gain} of each modality with respect to the task prompt. Let $\mathbf{p}$ denote the embedding of the input prompt (e.g., user query), and $\mathbf{z}_m$ denote the compressed representation of modality $m$ extracted from the early layers of the corresponding encoder. We compute the conditional relevance score $\alpha_m$ as:

\begin{equation}
\alpha_m = \text{MLP}\left( [\mathbf{p}; \mathbf{z}_m] \right) \in \mathbb{R},
\end{equation}
where $[\cdot;\cdot]$ denotes concatenation. The scores are normalized across modalities using a softmax function: $\beta_m = \frac{\exp(\alpha_m)}{\sum_{m' \in \mathcal{M}} \exp(\alpha_{m'})}$, where $\mathcal{M}$ is the set of input modalities.

\subsubsection{Modal Activation Sparsity (MAS)}

Based on the above multidimensional analysis, we propose a unified metric, \textit{Modal Activation Sparsity} (MAS), to quantify the overall necessity of each modality in the current inference context. The MAS for modality $m$ is defined as:

\begin{equation}
\text{MAS}_m = 1 - \beta_m \cdot \left(1 - \lambda_{\text{spatial}} \cdot \rho_{\text{spatial}}^{(m)} - \lambda_{\text{temp}} \cdot \gamma_{\text{avg}}^{(m)}\right),
\end{equation}
where $\rho_{\text{spatial}}^{(m)}$ and $\gamma_{\text{avg}}^{(m)}$ are the spatial and temporal sparsity measures specific to modality $m$ (applicable only when the modality exhibits such dimensions), and $\lambda_{\text{spatial}}, \lambda_{\text{temp}} \in [0,1]$ are hyperparameters controlling the contribution of spatial and temporal redundancy. A high $\text{MAS}_m$ indicates that modality $m$ contains substantial redundancy or low task relevance, suggesting that it can be aggressively compressed or even omitted during inference. Conversely, a low $\text{MAS}_m$ signals critical information that must be preserved with high fidelity. This lightweight probing module introduces minimal computational overhead, only requiring forward passes through early encoder layers and a few lightweight heads, yet provides fine-grained guidance for the subsequent edge-cloud offloading strategy. The computed MAS values, together with the spatial and temporal sparsity maps, serve as compact descriptors of the input's computational characteristics, enabling adaptive resource allocation between edge and cloud.

\begin{figure}[t]
  \centering
 \includegraphics[width=1\columnwidth]{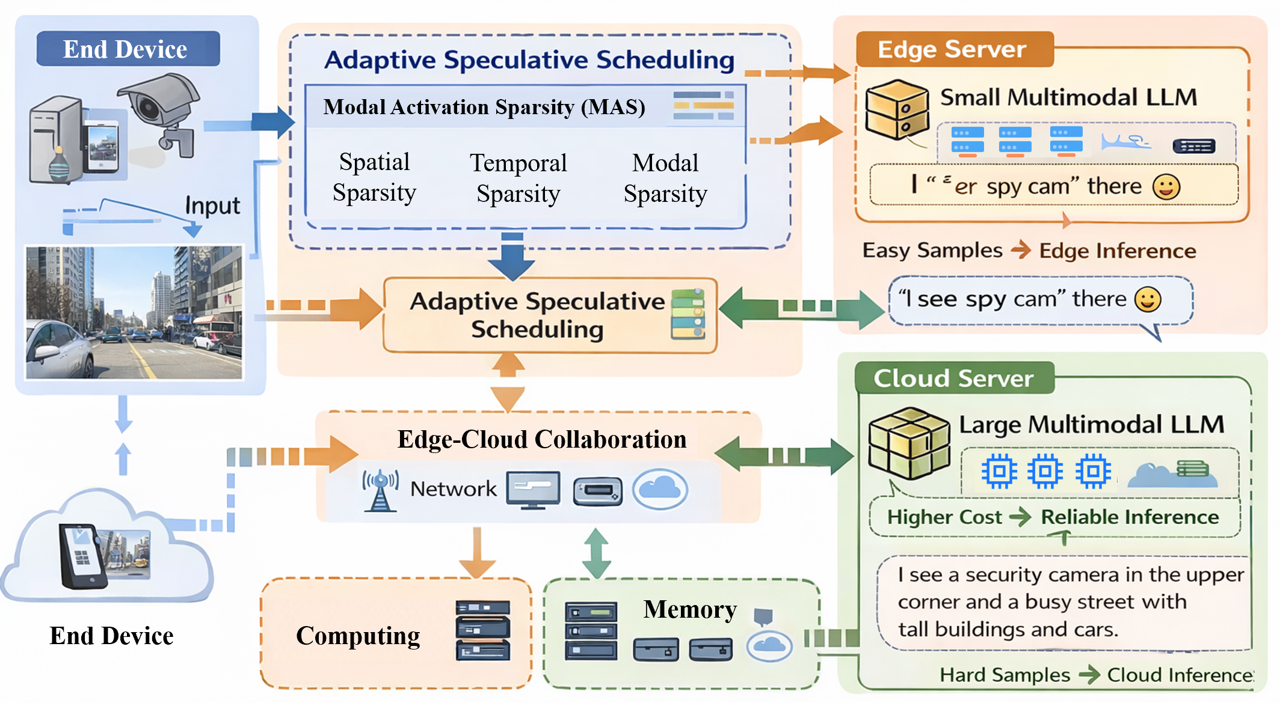}
  \caption{The illustration of adaptive speculative edge-cloud collaborative offloading.}
  \label{figure3}
  \vspace{-2mm}
\end{figure}

\subsection{Adaptive Speculative Edge-Cloud Collaborative Offloading}
Building upon the Modal Activation Sparsity (MAS) metrics derived from the lightweight probing module, we now address the core challenge of optimal offloading decisions in heterogeneous edge-cloud environments. Unlike conventional deterministic offloading strategies, we propose an adaptive speculative edge-cloud collaborative offloading mechanism, as illustrated in Figure~\ref{figure3}.  It leverages the inherent uncertainty in MLLM inference to minimize end-to-end latency while maintaining generation quality.

\subsubsection{Problem Formulation}

Consider an MLLM inference task with $K$ input modalities, each characterized by its MAS value $\text{MAS}_m \in [0,1]$ obtained from the previous stage. The inference process consists of two phases: (1) \textit{prefill phase}, where input tokens are processed to generate the key-value (KV) cache, and (2) \textit{decoding phase}, where tokens are generated autoregressively. We introduce a \textit{speculative execution} framework where the edge device maintains a lightweight draft model (a subset of the MLLM layers) capable of generating \textit{draft tokens} rapidly. The cloud server hosts the full MLLM for verification and correction. The key innovation lies in adaptively determining: (i) which modalities to offload, (ii) when to invoke speculative execution, and (iii) the confidence threshold that triggers asynchronous offloading.

Let $D_{\text{edge}}^{\text{prefill}}(\boldsymbol{\beta}, \boldsymbol{\rho})$ denote the prefill latency on the edge, which depends on the retained modality proportions $\boldsymbol{\beta} = [\beta_1, \ldots, \beta_K]$ and spatial-temporal compression ratios $\boldsymbol{\rho} = [\rho_1, \ldots, \rho_K]$. Similarly, $D_{\text{cloud}}^{\text{prefill}}(\boldsymbol{\beta}, \boldsymbol{\rho})$ represents cloud prefill latency. The communication latency for transmitting modality $m$ is:

\begin{equation}
T_{\text{comm}}^{(m)} = \frac{\text{DataSize}(\boldsymbol{\beta}_m, \boldsymbol{\rho}_m)}{B_{\text{eff}}} + RTT,
\end{equation}
where $B_{\text{eff}}$ is the effective bandwidth and $RTT$ is the round-trip time. For the decoding phase, we model speculative generation as follows. The edge draft model generates $N_{\text{draft}}$ tokens with latency $T_{\text{draft}} = N_{\text{draft}} \cdot t_{\text{draft}}$, where $t_{\text{draft}}$ is per-token latency on the edge. The cloud verification latency for these $N_{\text{draft}}$ tokens is $T_{\text{verify}} = N_{\text{draft}} \cdot t_{\text{verify}}^{\text{cloud}}$. The acceptance probability of draft tokens is governed by the model's confidence distribution.

Then, we define the confidence of the draft model at generation step $i$ as the entropy of the output probability distribution:

\begin{equation}
\mathcal{H}(\mathbf{p}_i) = -\sum_{v \in \mathcal{V}} p_i(v) \log p_i(v),
\end{equation}
where $\mathcal{V}$ is the vocabulary. A lower entropy indicates higher confidence. We introduce a \textit{confidence threshold} $\theta_{\text{conf}} \in [0,1]$ such that:

\begin{equation}
\text{Speculate}(i) = 
\begin{cases}
\text{True}, & \text{if } \mathcal{H}(\mathbf{p}_i) \leq \theta_{\text{conf}}. \\
\text{False}, & \text{otherwise}.
\end{cases}
\end{equation}

When $\mathcal{H}(\mathbf{p}_i) > \theta_{\text{conf}}$, the edge immediately offloads the intermediate state to the cloud for high-quality generation. Our goal is to minimize the expected end-to-end latency $L_{\text{e2e}}$ while ensuring that the generation quality degradation $\Delta Q$ remains bounded. The decision variables include:

\begin{itemize}
    \item $\boldsymbol{\beta} \in [0,1]^K$: modality retention ratios
    \item $\boldsymbol{\rho} \in [0,1]^K$: compression ratios per modality
    \item $\theta_{\text{conf}} \in [0,1]$: confidence threshold
    \item $N_{\text{draft}} \in \mathbb{N}^+$: speculative decoding length
\end{itemize}

The optimization problem is formulated as:

\begin{equation}
\begin{aligned}
& \underset{\boldsymbol{\beta}, \boldsymbol{\rho}, \theta_{\text{conf}}, N_{\text{draft}}}{\text{minimize}} & & \mathbb{E}\left[ L_{\text{e2e}}(\boldsymbol{\beta}, \boldsymbol{\rho}, \theta_{\text{conf}}, N_{\text{draft}}) \right] \\
& \text{subject to} & & \Delta Q(\boldsymbol{\beta}, \boldsymbol{\rho}) \leq \epsilon_Q, \\
& & & \sum_{m=1}^{K} \text{Mem}_{\text{edge}}(\beta_m, \rho_m) \leq \text{Mem}_{\text{edge}}^{\max}, \\
& & & T_{\text{comm}}^{(m)} \leq T_{\text{max}}^{(m)}, \quad \forall m \in \mathcal{M}, \\
& & & \beta_m \geq 1 - \text{MAS}_m, \quad \forall m \in \mathcal{M},
\end{aligned}
\end{equation}
where $\epsilon_Q$ is the maximum tolerable quality degradation, $\text{Mem}_{\text{edge}}$ is edge memory consumption, and the final constraint ensures that modalities with low MAS (high importance) cannot be aggressively compressed.

The expected end-to-end latency comprises three components: prefill latency, speculative decoding latency, and offloading latency for low-confidence steps. Let $P_{\text{conf}}(\theta_{\text{conf}})$ be the probability that the model's confidence exceeds the threshold at a given step, approximated by the empirical cumulative distribution of entropy values:

\begin{equation}
P_{\text{conf}}(\theta_{\text{conf}}) = \frac{1}{N_{\text{steps}}} \sum_{i=1}^{N_{\text{steps}}} \mathbb{1}\left[ \mathcal{H}(\mathbf{p}_i) \leq \theta_{\text{conf}} \right].
\end{equation}

The expected number of tokens generated via speculative execution is:

\begin{equation}
\mathbb{E}[N_{\text{spec}}] = \sum_{t=1}^{\infty} t \cdot P_{\text{conf}}(\theta_{\text{conf}})^{t-1} (1 - P_{\text{conf}}(\theta_{\text{conf}})) = \frac{1}{1 - P_{\text{conf}}(\theta_{\text{conf}})}.
\end{equation}

The total expected latency is then:

\begin{equation}
\begin{aligned}
\mathbb{E}[L_{\text{e2e}}] = &\; \max\left(D_{\text{edge}}^{\text{prefill}}(\boldsymbol{\beta}, \boldsymbol{\rho}),\; D_{\text{cloud}}^{\text{prefill}}(\boldsymbol{\beta}, \boldsymbol{\rho}) + \sum_{m} \mathbb{1}_{\text{offload}}^{(m)} T_{\text{comm}}^{(m)}\right) \\
& + \mathbb{E}[N_{\text{spec}}] \cdot \left( T_{\text{draft}} + P_{\text{conf}}(\theta_{\text{conf}}) \cdot T_{\text{verify}} \right) \\
& + \left(1 - P_{\text{conf}}(\theta_{\text{conf}})\right) \cdot T_{\text{offload}},
\end{aligned}
\end{equation}
where $\mathbb{1}_{\text{offload}}^{(m)}$ indicates whether modality $m$ requires offloading, and $T_{\text{offload}}$ is the latency for offloading intermediate states during low-confidence steps.

\subsubsection{Adaptive Speculative Edge-Cloud Collaborative Offloading Algorithm}

Directly solving the optimization problem in Eq.~(11) is challenging due to its non-convex objective and the strong coupling among decision variables $\boldsymbol{\beta}$, $\boldsymbol{\rho}$, $\theta_{\text{conf}}$, and $N_{\text{draft}}$. To address this challenge, we propose an adaptive speculative edge-cloud collaborative offloading algorithm that iteratively optimizes these variables based on real-time system states, as presented in Algorithm 1. The algorithm operates on two distinct timescales to balance global optimality with runtime adaptability:

\begin{itemize}
    \item \textbf{Coarse-grained optimization (per-request level):} Executed once per inference request, this phase determines the modality retention ratios $\boldsymbol{\beta}$, compression ratios $\boldsymbol{\rho}$, and initial offloading decisions using the MAS metrics derived from the lightweight probing module. These decisions remain fixed throughout the decoding process for a given request.
    
    \item \textbf{Fine-grained adaptation (per-step level):} Performed at each decoding step, this phase dynamically adjusts the confidence threshold $\theta_{\text{conf}}$ based on observed model confidence and triggers speculative execution when appropriate. This fast-timescale adaptation enables the system to respond to varying input characteristics and real-time network conditions.
\end{itemize}

\begin{algorithm}
\caption{Adaptive Speculative Edge-Cloud Collaborative Offloading Algorithm}
\label{alg:aseco}
\begin{algorithmic}[1]
\Require MAS values $\{\text{MAS}_m\}$, system state $\mathcal{S}$, quality bound $\epsilon_Q$
\Ensure Offloading decisions $\boldsymbol{\beta}^*, \boldsymbol{\rho}^*, \theta_{\text{conf}}^*, N_{\text{draft}}^*$

\State $\boldsymbol{\beta}^*, \boldsymbol{\rho}^* \gets \arg\min \mathbb{E}[L_{\text{prefill}}]$ s.t. $\Delta Q \leq \epsilon_Q,\ \beta_m \geq 1-\text{MAS}_m$ 
\State $\theta_{\text{conf}}^* \gets \mathcal{H}_{\text{emp}}^{-1}(0.7)$
\State $N_{\text{draft}}^* \gets \min\left(\left\lfloor \frac{\log(1-P_{\text{target}})}{\log P_{\text{conf}}}\right\rfloor, N_{\text{max}}\right)$

\For{each decoding step $i$}
    \State Compute draft token $y_i^{\text{draft}}$ and confidence $\mathcal{H}(\mathbf{p}_i)$
    \If{$\mathcal{H}(\mathbf{p}_i) \leq \theta_{\text{conf}}^*$ and $\text{cache} \geq N_{\text{draft}}^*$}
        \State Send cache to cloud for parallel verification
        \State Update $\theta_{\text{conf}}^*$ with EMA of accepted tokens
    \ElsIf{$\mathcal{H}(\mathbf{p}_i) > \theta_{\text{conf}}^*$}
        \State Asynchronously offload intermediate state $\mathcal{I}_i$ to cloud
        \State $\theta_{\text{conf}}^* \gets \max(\theta_{\text{conf}}^* \cdot \delta,\ \theta_{\text{min}})$
    \EndIf
\EndFor

\State \textbf{return} $\boldsymbol{\beta}^*, \boldsymbol{\rho}^*, \theta_{\text{conf}}^*, N_{\text{draft}}^*$
\end{algorithmic}
\end{algorithm}

The algorithm exhibits several desirable properties. For the coarse-grained optimization phase, we employ Bayesian optimization with Gaussian process surrogate models, which achieves sublinear regret:

\begin{equation}
\mathbb{E}[R_T] = O\left(\sqrt{T \cdot \log T}\right),
\end{equation}
where $T$ is the number of function evaluations and $R_T$ is the cumulative regret relative to the global optimum. For the fine-grained adaptation, the exponential moving average (EMA) update ensures that $\theta_{\text{conf}}^*$ converges to the optimal threshold that balances speculation success rate and offloading overhead:

\begin{equation}
\lim_{n \to \infty} \mathbb{E}[|\theta_{\text{conf}}^{(n)} - \theta_{\text{conf}}^*|^2] = 0.
\end{equation}

\paragraph{Complexity Analysis.}
The computational complexity of it is $O(K \cdot \log K + N_{\text{steps}} \cdot M_{\text{draft}})$, where $M_{\text{draft}}$ is the draft model size, which is significantly smaller than the full MLLM complexity $O(N_{\text{steps}} \cdot M_{\text{full}})$, achieving an acceleration factor of approximately $M_{\text{full}} / M_{\text{draft}}$ for high-confidence steps.

\section{Experiments}
\label{sec:experiments}

\subsection{Experimental Setup}

\subsubsection{Datasets and Implementation Details}
 We evaluate our method on two widely adopted multimodal benchmarks: \textbf{(1) VQAv2} \cite{goyal2017making161616}: A large-scale visual question answering dataset containing over 250,000 images and 1.1 million questions. We randomly sample 5,000 images from the validation set to construct our test set, ensuring diverse coverage of question types and visual content. \textbf{(2) MMBench} \cite{liu2024mmbench171717}: A comprehensive multimodal benchmark that evaluates model performance across 20 distinct capability dimensions, including object recognition, attribute reasoning, and scene understanding. We utilize the full test set for evaluation.

We conduct experiments on a heterogeneous edge-cloud platform. The cloud server is equipped with an NVIDIA A100 (40 GB) GPU, while the edge device is configured with an NVIDIA RTX 3090 (24 GB) GPU. 
We deploy a lightweight draft model on the edge device and a full-capacity model on the cloud. Specifically, we adopt Qwen2-VL-2B \cite{bai2023qwen181818} as the edge-side draft model, which features 2 billion parameters and is capable of rapid token generation. The cloud-side model is Qwen2.5-VL-7B \cite{bai2023qwen181818}, a 7-billion-parameter multimodal large language model that delivers high-quality generation. The two models share the same tokenizer and architectural design, enabling seamless speculative verification. To simulate realistic edge-cloud communication scenarios, we configure the network bandwidth at three representative levels: 200 Mbps (low-bandwidth), 300 Mbps (medium-bandwidth), and 400 Mbps (high-bandwidth). The round-trip time (RTT) is fixed at 20 ms to reflect typical edge-cloud latency. All reported results are averaged over three independent runs to ensure statistical reliability.

\subsubsection{Baseline Methods}
We compare our proposed MSAO framework against the following baseline approaches. \textbf{(1) Cloud-only}: The full Qwen2.5-VL-7B model runs entirely on the cloud server. All multimodal inputs are transmitted to the cloud for inference. \textbf{(2) Edge-only}: The lightweight Qwen2-VL-2B model runs entirely on the edge device. \textbf{(3) PerLLM} \cite{yang2024perllm101010}: An edge-cloud collaborative inference framework that partitions MLLM execution based on layer-wise offloading. 

\subsubsection{Evaluation Metrics}

We evaluate the proposed method using the following metrics:

\begin{itemize}
    \item \textbf{Accuracy}: For VQAv2, we report the standard VQA accuracy, which measures the exact match between predicted answers and ground-truth annotations. For MMBench, we report the average accuracy across all 20 capability dimensions. These metrics assess the generation quality preservation under edge-cloud collaboration.

    \item \textbf{Throughput}: The number of tokens generated per second (tokens/s), which reflects the system's processing efficiency under continuous inference workloads.
    
    \item \textbf{End-to-End Latency}: The total time from when the input is submitted to when the final response is fully generated. This includes prefill phase latency, decoding phase latency, and all communication overheads. We report the average latency per inference request (in milliseconds).
    
    \item \textbf{Computing Overhead}: This metric quantifies the total floating-point operations (FLOPs) consumed during inference, capturing the computational burden on both edge and cloud devices. 

    \item \textbf{Memory Overhead}: This metric measures the peak GPU memory consumption during inference, including model weights, activations, and KV cache occupancy. 
    
\end{itemize}

\subsubsection{Parameter Configuration}
The spatial sparsity threshold $\tau_s$ is set to 0.3 based on empirical sensitivity analysis. The temporal redundancy weight $\lambda_{\text{temp}}$ is set to 0.4, and the spatial redundancy weight $\lambda_{\text{spatial}}$ is set to 0.6, reflecting the observation that spatial redundancy often contributes more to overall computational savings than temporal redundancy in image-based tasks. The maximum tolerable quality degradation $\epsilon_Q$ is set to 2\% relative to the Cloud-only baseline, ensuring that the proposed method maintains generation fidelity. The initial confidence threshold $\theta_{\text{conf}}^*$ is set to the 70th percentile of the empirical entropy distribution collected from a calibration set of 500 samples. The threshold decay factor $\delta$ is set to 0.95, balancing adaptation speed and stability. The maximum speculative length $N_{\text{max}}$ is set to 5, consistent with prior work on speculative decoding. For coarse-grained optimization, we perform 50 iterations of Bayesian optimization using a Gaussian process surrogate with Matérn 5/2 kernel. The acquisition function is expected to improve with an exploration-exploitation trade-off parameter of 0.1. The target acceptance probability $P_{\text{target}}$ is set to 0.8, providing a balance between speculation aggressiveness and verification overhead.


\begin{figure}[t]
  \centering
 \includegraphics[width=1\columnwidth]{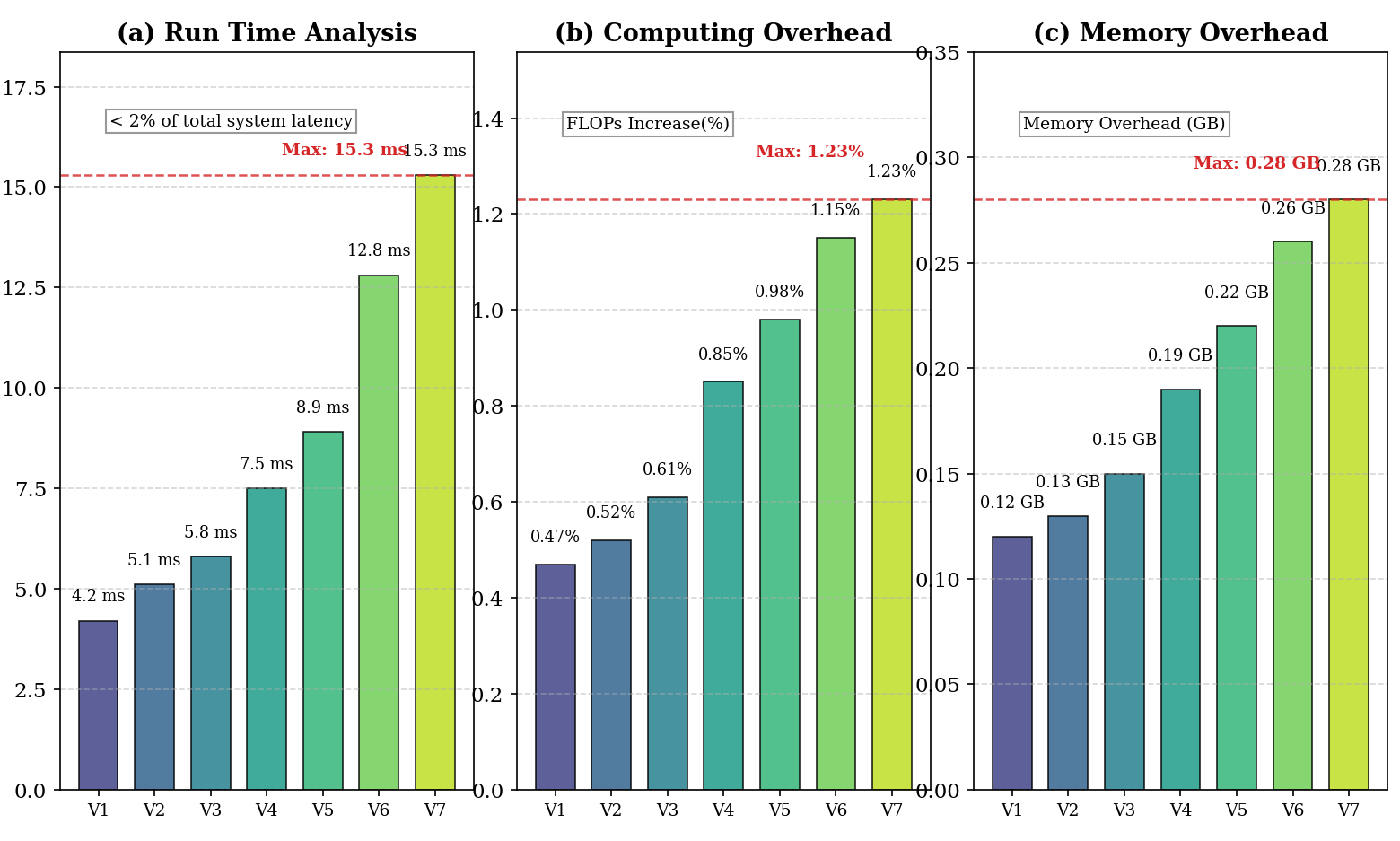}
  \caption{The performance analysis of lightweight heterogeneous modality-aware processing.}
  \label{figure4}
  \vspace{-2mm}
\end{figure}

\subsection{Lightweight Heterogeneous Modality-Aware Analysis}
To validate the lightweight nature of our proposed modality-aware module, we conduct a comprehensive analysis of its computational characteristics under varying configurations. Specifically, we report results for seven representative configurations (V1–V7), covering unimodal text, bimodal image-text, and trimodal video-text-audio inputs across increasing resolution and sequence length.

As illustrated in Figure~\ref{figure4}, the average inference latency contributed by the aware module ranges from 4.2 ms to 15.3 ms, depending on input complexity, which constitutes less than 2\% of the total end-to-end inference latency in typical edge-cloud deployment scenarios. Notably, even under the most demanding configuration, the module completes its analysis within 15.3 ms, ensuring that it does not become a bottleneck in the inference pipeline. In terms of computing overhead, the module introduces a minimal FLOPs increase of only 0.47\% to 1.23\% relative to the full MLLM inference pipeline, demonstrating its highly efficient design. Specifically, the spatial aware head accounts for the majority of this overhead, while the temporal and modal aware components contribute negligible additional computation due to their reliance on lightweight hash operations and MLP projections.  In terms of memory overhead, the aware module adds only 0.12 GB to 0.28 GB of additional GPU memory footprint on the edge device. This marginal increase is primarily attributed to the storage of intermediate feature maps from the early encoder layers, which are already resident in memory for subsequent processing. The spatial importance maps and temporal similarity caches are stored as lightweight tensors, contributing negligible memory pressure.  Collectively, these results confirm that the proposed lightweight modality-aware module achieves its design objective of providing fine-grained sparsity analysis without imposing significant computational, memory, or latency burdens on the overall system.

\subsection{Main Results and Analysis}
\label{subsec:main_result}

\subsubsection{Accuracy Comparison}
As shown in Table~\ref{tab1}, the proposed MSAO framework achieves accuracy comparable to the cloud-only upper bound while significantly outperforming edge-only and PerLLM baselines. Specifically, MSAO maintains a minimal performance gap of less than 0.4\% relative to the cloud-only approach across both datasets and various bandwidth conditions (e.g., 76.1\% vs. 76.3\% on VQAv2 at 200 Mbps, and 76.3\% vs. 76.5\% on MMBench at 400 Mbps). In contrast, MSAO substantially surpasses the edge-only and PerLLM methods by large margins, with absolute accuracy improvements ranging from 4.8\% to 16.8\% across all settings. These results demonstrate that MSAO effectively preserves the high accuracy of cloud-based inference through adaptive heterogeneous offloading, while achieving substantial gains over purely edge-based or suboptimal adaptive strategies.


\begin{table}[h]
\centering
\caption{Accuracy (\%) comparison results.}
\label{tab1}
\resizebox{\linewidth}{!}{
\begin{tabular}{|l|c|c|c|c|}
\hline
\multicolumn{1}{|c|}{\textbf{Method}} & \textbf{Cloud-only} & \textbf{Edge-only} & \textbf{PerLLM} & \textbf{MSAO} \\ \hline
\multicolumn{5}{|c|}{\textit{VQAv2 Dataset}} \\ \hline
200 (Mbps) & 76.3 & 61.4 & 71.3 & 76.1\\ \hline
300 (Mbps) & 77.4 & 63.2 & 71.8 & 77.2\\ \hline
400 (Mbps)   & 77.8 & 63.5 & 72.4  & 77.5\\ \hline
\multicolumn{5}{|c|}{\textit{MMBench Dataset}} \\ \hline
200 (Mbps) & 75.6 & 58.4 & 68.3 & 75.2\\ \hline
300 (Mbps) & 76.1 & 60.1 & 69.2 & 75.9\\ \hline
400 (Mbps)   & 76.5 & 61.2 & 69.9  & 76.3\\ \hline
\end{tabular}
}
\end{table}

\subsubsection{Throughput Comparison}
\begin{figure}[t]
  \centering
 \includegraphics[width=1\columnwidth]{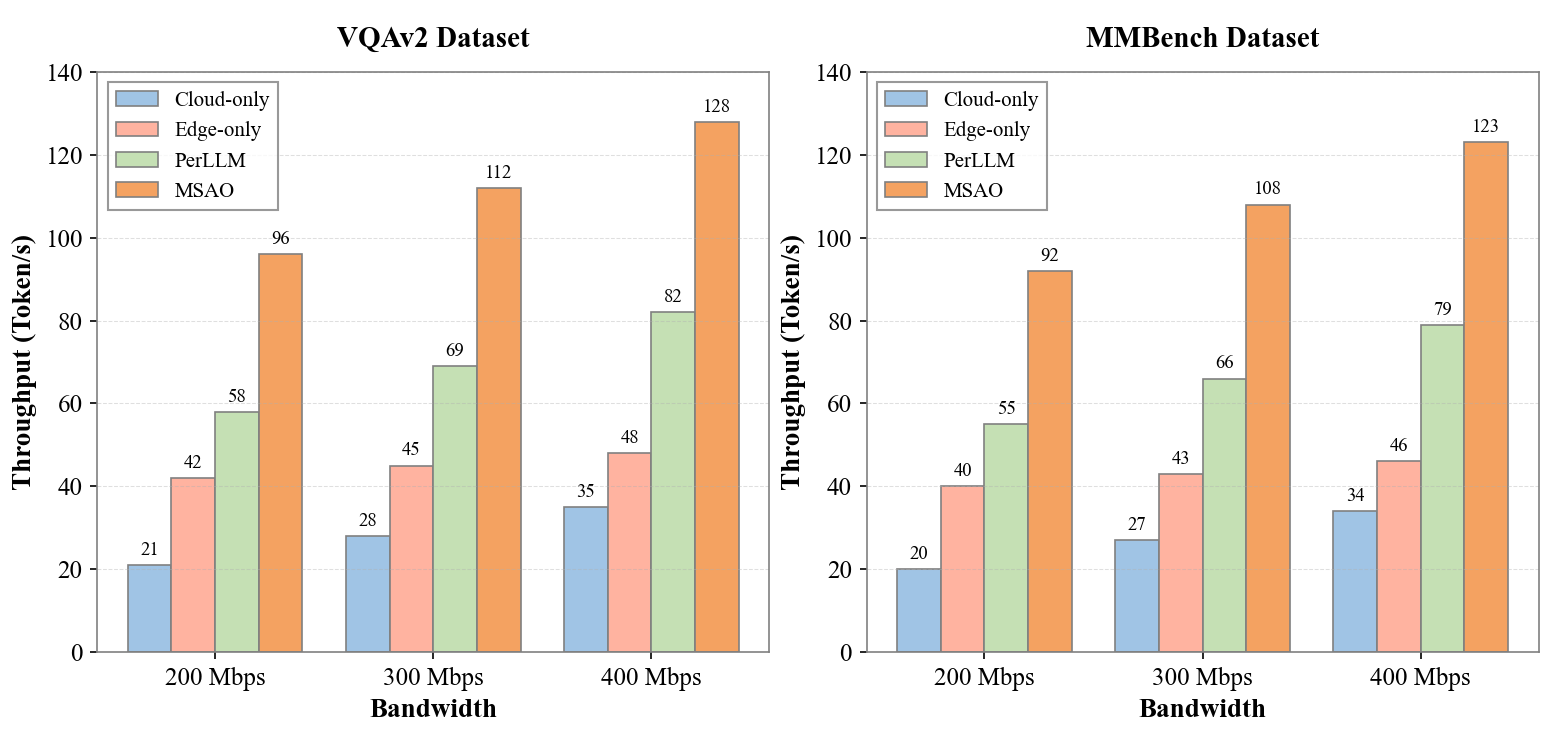}
  \caption{The throughput comparison results of different methods.}
  \label{figure5}
\end{figure}

We evaluate the system throughput of different methods under varying bandwidth conditions (200, 300, and 400 Mbps) on the VQAv2 and MMBench datasets. As illustrated in Figure~\ref{figure5}, the proposed MSAO framework consistently achieves the highest throughput across all settings. Specifically, MSAO attains throughput improvements ranging from 1.5$\times$ to 2.3$\times$ compared to the Cloud-only baseline, which is constrained by heavy uplink transmission and cloud inference latency. For instance, on the VQAv2 dataset at 400 Mbps, MSAO achieves 128 Token/s, while Cloud-only reaches only 35 Token/s, representing a 2.66$\times$ improvement. Compared to the Edge-only baseline, which is limited by the lightweight model's capacity, MSAO achieves over 2$\times$ higher throughput across all bandwidth settings. Furthermore, MSAO consistently outperforms PerLLM by a substantial margin, with improvements ranging from 1.5$\times$ to 1.7$\times$. These results demonstrate that MSAO's adaptive modality sparsity-aware offloading strategy effectively balances edge-cloud workloads, enabling efficient parallel processing and significantly reducing communication overhead, thereby achieving superior throughput efficiency.

\subsubsection{End-to-End Latency Comparison}
\begin{figure}[th]
	\centering 
    \subfigure[VQAv2]{
    \label{Fig.sub.3.1}
    \includegraphics[width=0.22\textwidth]{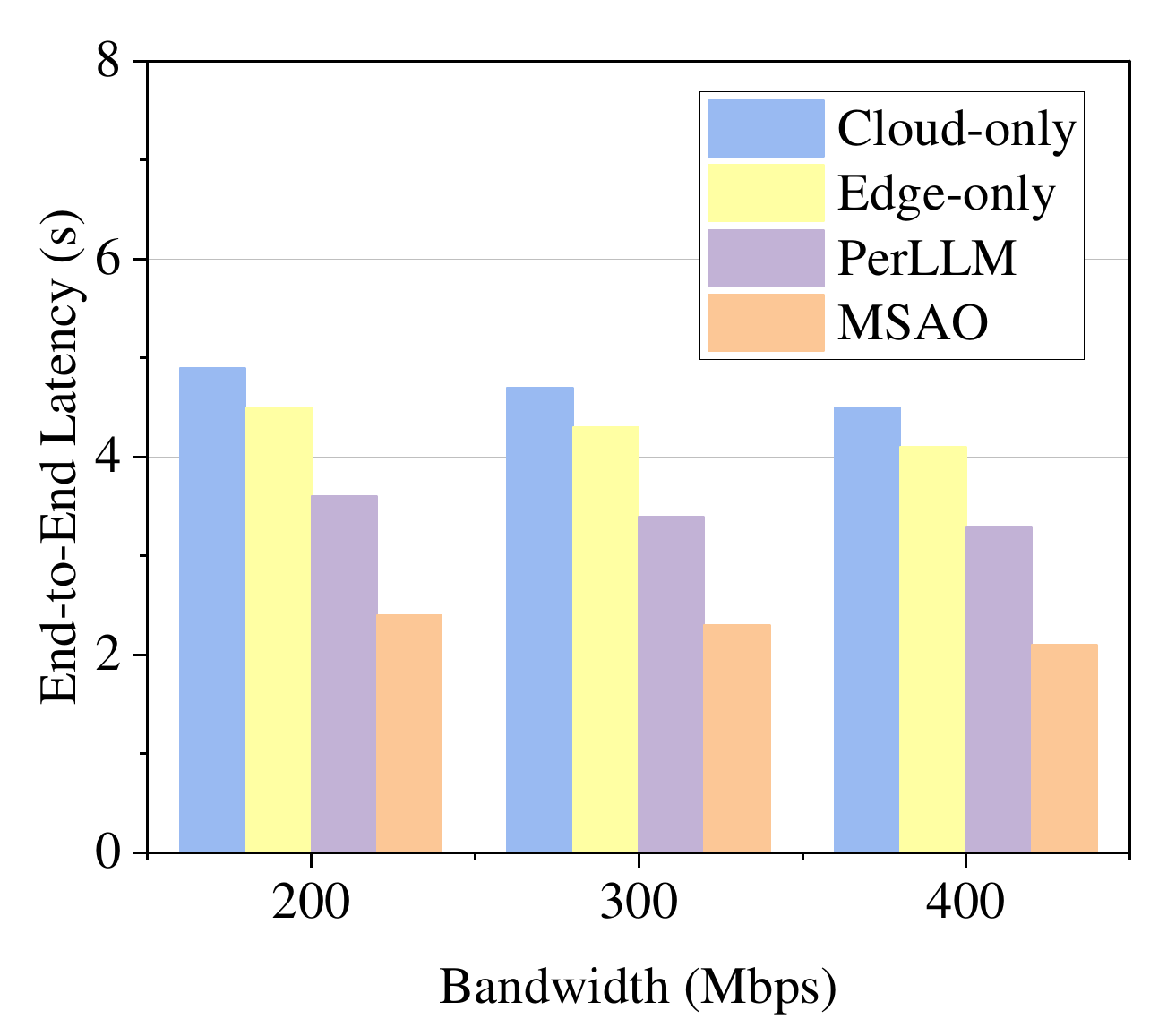}}
    \subfigure[MMBench ]{
    \label{Fig.sub.3.2}
    \includegraphics[width=0.22\textwidth]{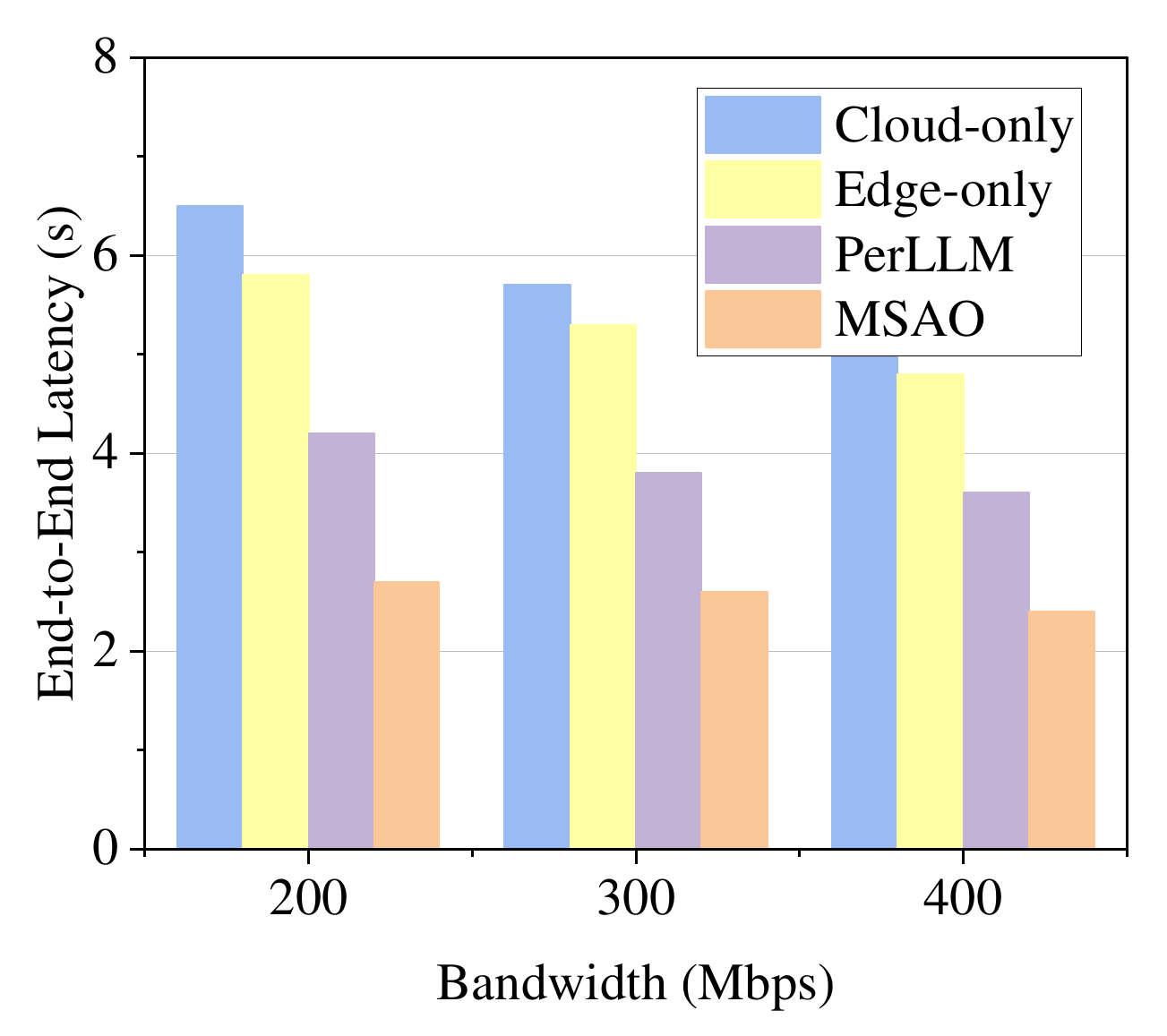}}
    \caption{The end-to-end latency comparison results of different methods.}
	\label{figure6}
\end{figure}

As shown in Figure~\ref{figure6}, the proposed MSAO framework achieves the lowest mean end-to-end latency across all evaluated strategies under varying bandwidth conditions. Specifically, MSAO reduces latency by over 50\% compared to the Cloud-only approach, which suffers from heavy uplink transmission and cloud inference overhead. Similarly, it achieves a latency reduction of approximately 45\%–55\% relative to the Edge-only approach, which is constrained by the limited computational capacity of edge devices and exhibits pronounced latency tails when processing complex multimodal inputs. Furthermore, compared to PerLLM, MSAO achieves over 30\% latency reduction, demonstrating the superiority of its modality sparsity-aware adaptive offloading mechanism. By intelligently offloading only the most inference-intensive components to the cloud while retaining lightweight processing on the edge, MSAO effectively mitigates the latency bottlenecks inherent in both purely cloud-based and purely edge-based strategies. These results confirm that MSAO strikes an optimal balance between edge and cloud execution, achieving consistently low latency across diverse bandwidth conditions.


\subsubsection{Resource Overhead Comparison}

\begin{figure}[th]
	\centering 
    \subfigure[VQAv2, Computing]{
    \label{Fig.sub.4.1}
    \includegraphics[width=0.23\textwidth]{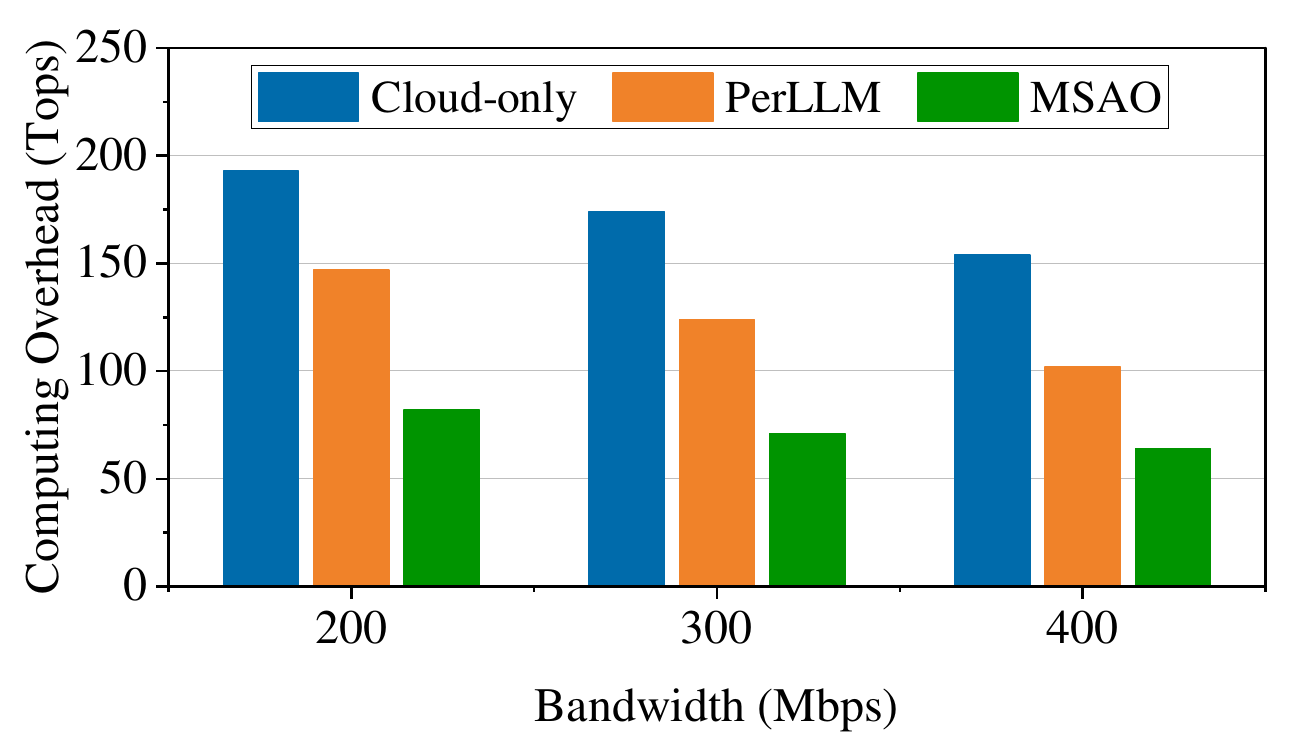}}
    \subfigure[MMBench, Computing   ]{
    \label{Fig.sub.4.2}
    \includegraphics[width=0.23\textwidth]{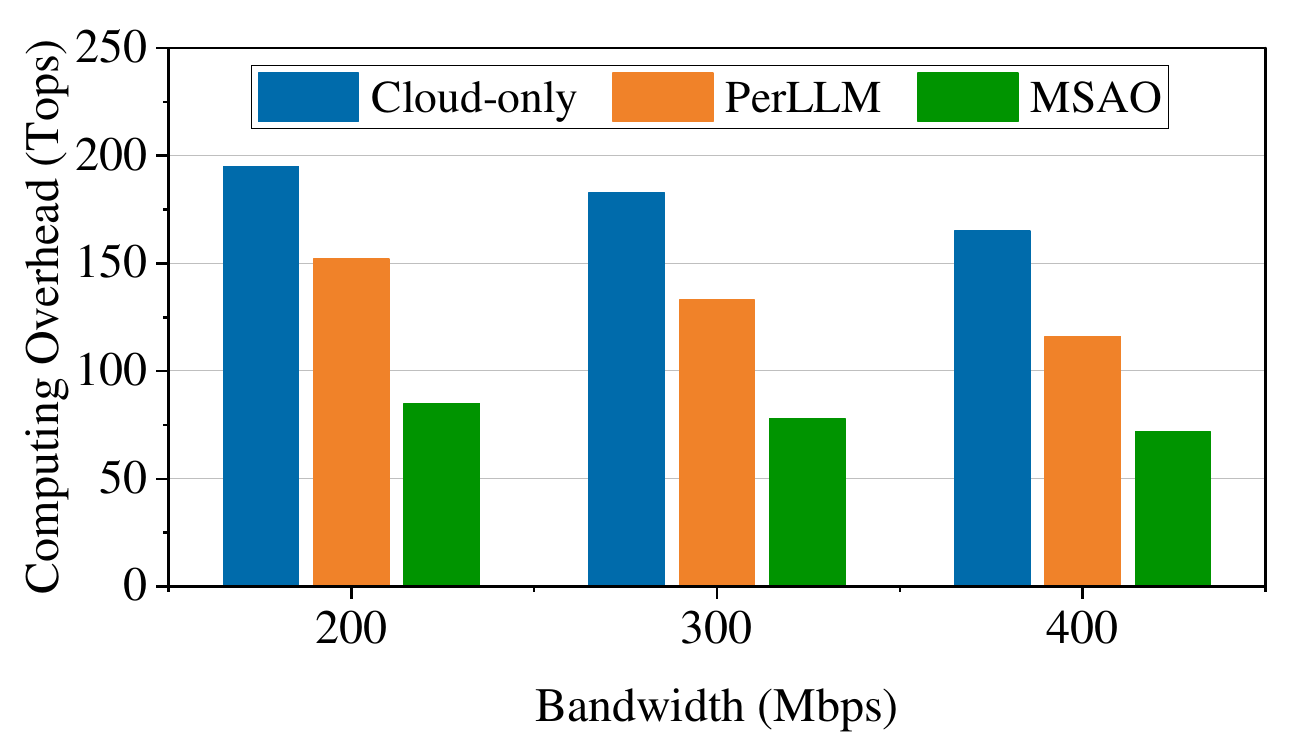}}
    \caption{The computing overhead comparison results of different methods.}
	\label{figure7}
\end{figure}

\textbf{Computing Overhead.} As shown in Figure~\ref{figure7}, the proposed MSAO framework achieves the lowest computing overhead on both the VQAv2 and MMBench datasets compared to all baselines. By adaptively offloading only the most complex multimodal samples to the cloud while processing simpler samples locally, MSAO substantially reduces the computational burden on cloud infrastructure. Specifically, MSAO achieves a computing overhead reduction of 30\%–65\% compared to the Cloud-only approach. Similarly, it outperforms PerLLM by reducing computing overhead by 35\%–50\%, demonstrating the superiority of MSAO's modality sparsity-aware selective offloading strategy. Notably, the overhead incurred on the edge device for computing the modality-aware offloading policy is negligible. This strategic workload distribution enables the system to resolve the vast majority of inference requests with minimal compute investment: simple tasks are efficiently handled on the edge, while only the most challenging ones are offloaded to the cloud, achieving an optimal balance between computational efficiency and inference quality.


\begin{figure}[th]
	\centering 
        \subfigure[VQAv2, Memory]{
    \label{Fig.sub.4.3}
    \includegraphics[width=0.23\textwidth]{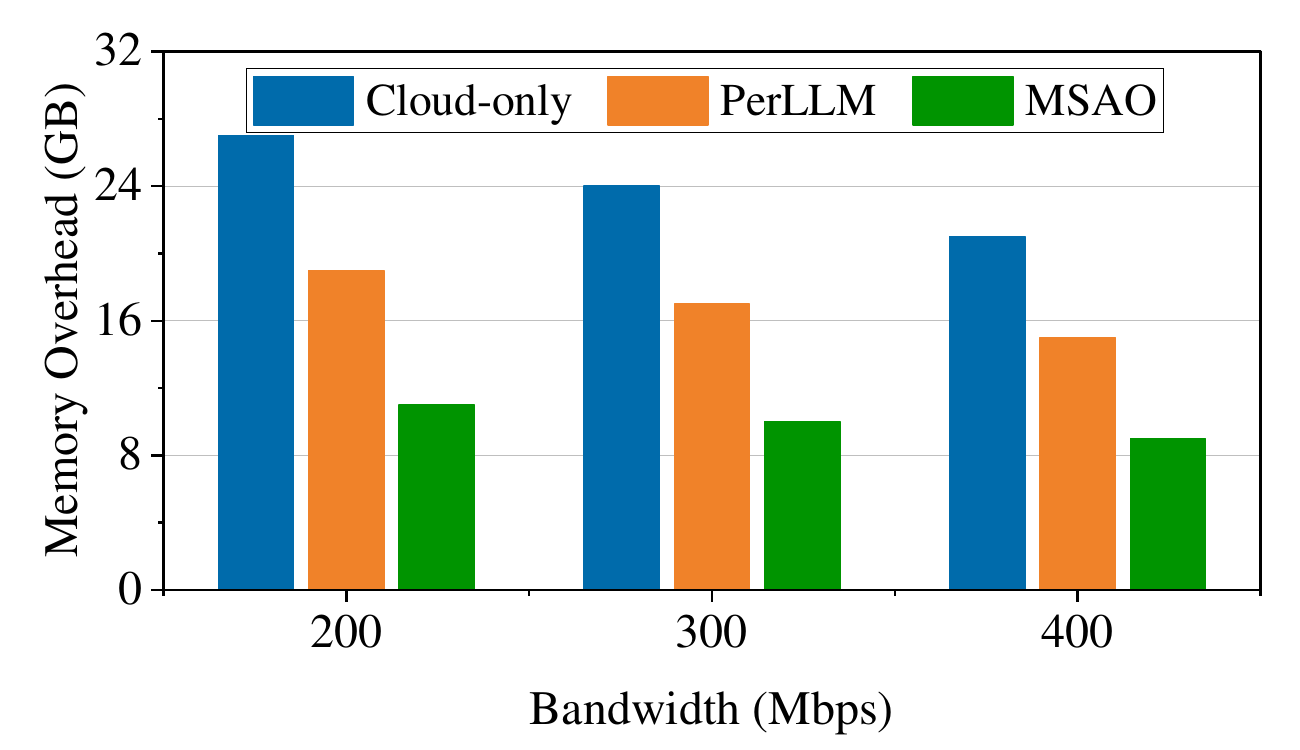}}
    \subfigure[MMBench, Memory]{
    \label{Fig.sub.4.4}
    \includegraphics[width=0.23\textwidth]{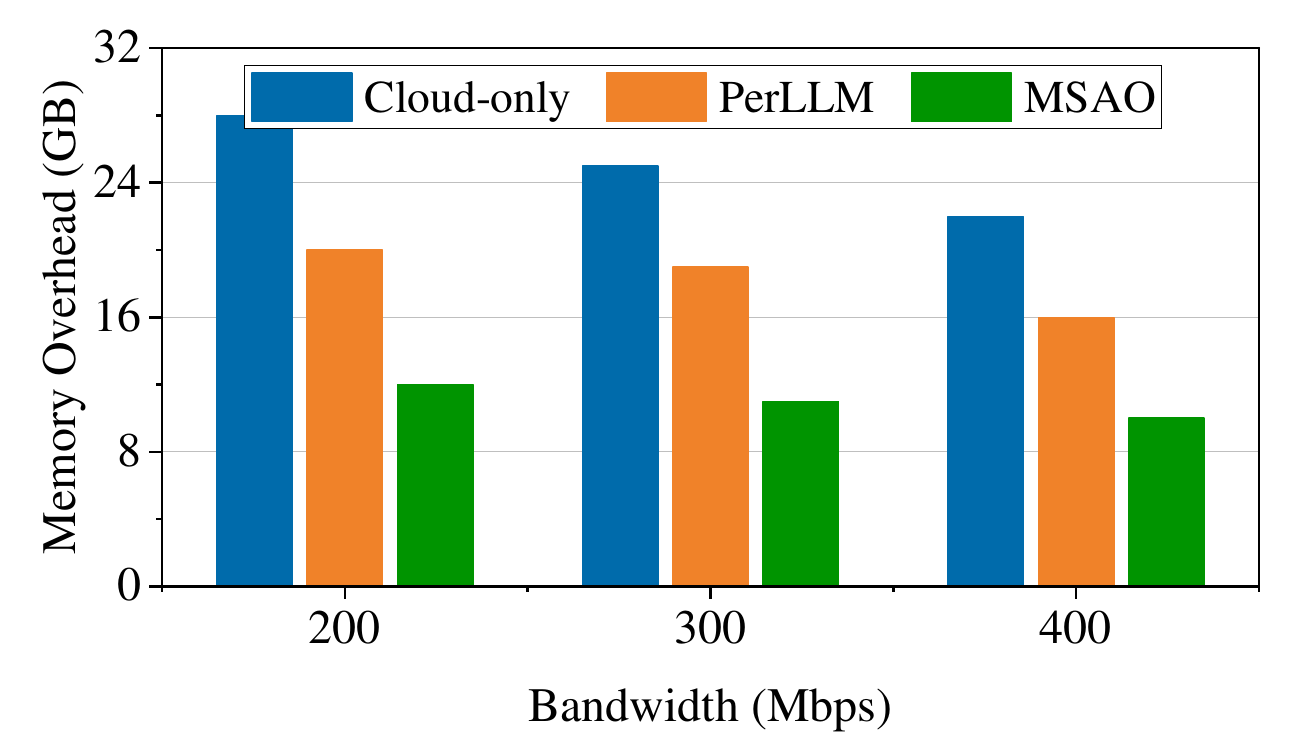}}  
    \caption{The memory overhead comparison results of different methods.}
	\label{figure8}
\end{figure}

\textbf{Memory Overhead.} As shown in Figure~\ref{figure8}, MSAO consistently achieves the lowest memory overhead across all bandwidth configurations. Specifically, under the 200 Mbps bandwidth setting, MSAO reduces memory overhead to 9.0 GB on the edge device, compared to 25.0 GB for Cloud-only and 17.0 GB for PerLLM, representing a 64\% reduction over Cloud-only and a 47\% reduction over PerLLM. Under the 300 Mbps configuration, MSAO achieves 8.5 GB, while Cloud-only and PerLLM consume 24.0 GB and 16.0 GB, respectively. Under the most favorable 400 Mbps setting, MSAO further reduces memory overhead to 8.5 GB, compared to 22.5 GB for Cloud-only and 15.5 GB for PerLLM. Overall, MSAO's lightweight sparsity-aware module identifies and prunes redundant modality information before offloading, while the speculative offloading mechanism ensures that only critical features are retained on the edge device. This design enables MSAO to achieve near-optimal memory efficiency across all bandwidth conditions, with negligible variation as bandwidth changes, confirming the robustness of our approach to network dynamics.


\subsection{Ablation Study}
To validate the effectiveness of key components in our MSAO framework, we conduct ablation studies by systematically removing two core modules: the adaptive modality-aware offloading mechanism and the edge-cloud collaborative scheduling strategy. Specifically, we compare the full MSAO framework against two variants: (1) w/o Modality-Aware, which disables the modality-aware offloading and instead uses a uniform offloading policy regardless of input modality complexity; and (2) w/o Collaborative Scheduling, which removes the edge-cloud collaborative scheduler and adopts a static task distribution strategy.

\begin{figure}[t]
  \centering
 \includegraphics[width=1\columnwidth]{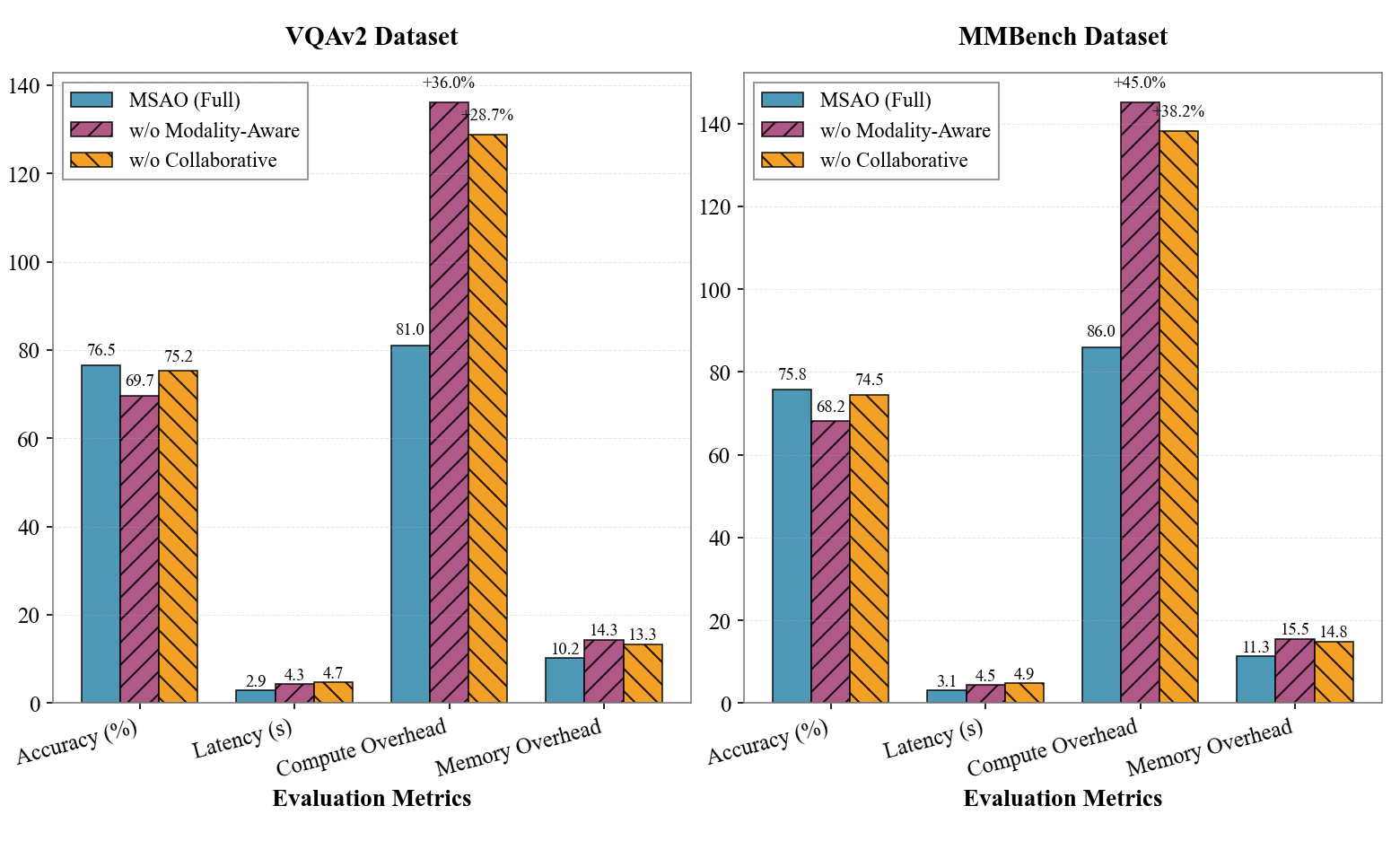}
  \caption{The ablation study results of the proposed MSAO framework.}
  \label{figure9}
\end{figure}

As shown in Figure~\ref{figure9}, disabling the modality-aware offloading mechanism leads to substantial accuracy degradation across both datasets, with accuracy dropping from 76.5\% to 69.7\% on VQAv2 and from 75.8\% to 68.2\% on MMBench, representing average declines of 6.8\% and 7.6\%, respectively. These results underscore the critical role of adaptively handling heterogeneous modality complexity. Furthermore, removing the collaborative scheduling strategy significantly degrades system efficiency. End-to-end latency increases from 2.9s to 4.3s on VQAv2 and from 3.1s to 4.5s on MMBench, corresponding to latency increases of 48.3\% and 45.2\%, respectively. Meanwhile, computational overhead rises from 4.3 to 4.7 on VQAv2 and from 4.5 to 4.9 on MMBench, while memory overhead increases from 2.9 to 4.3 on VQAv2 and from 3.1 to 4.5 on MMBench. These degradations stem from inefficient task distribution across edge and cloud, where static scheduling fails to adapt to dynamic bandwidth conditions and varying computational demands, leading to suboptimal resource utilization. Collectively, these results demonstrate that both adaptive modality-aware offloading and collaborative scheduling are indispensable for achieving efficient MLLM inference under resource-constrained edge-cloud environments.



\section{Conclusion}
In this work, we present MSAO, an adaptive edge-cloud collaborative inference framework for efficient MLLM inference. MSAO comprises two core components: (1) a lightweight fine-grained sparsity module that computes MAS via spatial-temporal-modal joint analysis, and (2) an adaptive speculative offloading mechanism that dynamically schedules edge-cloud workloads using confidence-guided execution. Extensive experiments on VQAv2 and MMBench demonstrate that MSAO reduces latency by 30\%, cuts resource overhead by 30\%--65\%, and improves throughput by 1.5$\times$--2.3$\times$ while maintaining competitive accuracy. Future directions include incorporating online adaptation for dynamic environments and evaluating on large-scale edge-cloud systems with practical deployment constraints.



\bibliographystyle{ACM-Reference-Format}
\bibliography{sample-base}

\appendix

\end{document}